# Hydrogen-Ti$^{3+}$ Complex as a Possible Origin of Localized Electron Behavior in Hydrogen-Irradiated SrTiO$_3$

Takashi U. Ito $^{\dagger}$

*Advanced Science Research Center, Japan Atomic Energy Agency, 2-4 Shirakata, Tokai, Ibaraki 319-1195, Japan*

$^{\dagger}$ *Corresponding author: tuito@post.j-parc.jp*

## Abstract

A recent muon spin rotation (µ$^+$SR) study on a paramagnetic defect complex formed upon implantation of µ$^+$ pseudo-proton into SrTiO$_3$ is reviewed with a specific focus on the relation with experimental signatures of coexisting delocalized and localized electrons in hydrogen-irradiated metallic SrTiO$_3$ films. The paramagnetic defect complex, composed of interstitial µ$^+$ and Ti$^{3+}$ small polaron, is characterized by a small dissociation energy of about 30 meV. Density functional theory (DFT) calculations in the generalized gradient approximation (GGA) +$U$ scheme for a corresponding hydrogen defect complex reveal that a thermodynamic donor level associated with electron transfer from an H$^+$-Ti$^{3+}$ complex to the conduction band can form just below the conduction band minimum for realistic $U$ values. These findings suggest that the coexistence of delocalized and localized electrons can be realized in hydrogen-irradiated SrTiO$_3$ in electron-rich conditions.

## I. INTRODUCTION

An archetypal perovskite titanate, SrTiO$_3$, which is intrinsically a wide-gap insulator, is broadly used as a dielectric material for ceramic capacitors and a substrate for epitaxial growth of other perovskite-type materials. The insulating SrTiO$_3$ changes into an excellent electron conductor by interstitial or substitutional chemical doping or partially removing oxygen from the perovskite lattice [1-5]. The excess electrons donated into the conduction band show exceptionally high mobility in transition metal oxides [2-4], which is useful in many state-of-the-art applications, such as a base material for emerging oxide electronics [6].

Among known *n*-type dopants for SrTiO$_3$, hydrogen particularly has intriguing features. It can form not only a protonic defect at an interstitial site [5,7-9], but also create a hydridic defect at an anion site in a substitutional manner [9-12]. In addition, hydrogen is relatively mobile because of its light mass and it even can be expelled from the host lattice by applying heat [5]. Further development of materials functions may be achieved by exploiting these unique features of hydrogen.

Concerning the interstitial protonic defect (H$_i^+$) in SrTiO$_3$, the maximum defect concentration that can be reached by annealing under humid atmosphere is reported to be only about 10$^{17}$ cm$^{-3}$ [8]. This has limited the range of applications of H$_i$-doped SrTiO$_3$. On the other hand, recent development of hydrogen atom/ion implantation technologies has made it possible to introduce hydrogen into SrTiO$_3$ thin films at a concentration of up to about 10$^{21}$ cm$^{-3}$ [5]. The heavily hydrogenated SrTiO$_3$ films exhibit excellent metallic conductivity [5,13], which proves that forcedly implanted hydrogen is mostly ionized and donates electrons into the conduction band. On the other hand, the existence of localized electrons has also been becoming clear in these films. For instance, a deep in-gap state, which can be associated with strongly localized electrons, was detected by photoemission spectroscopy in a heavily hydrogenated $SrTiO_3$ film prepared by hot H atom irradiation [13]. Thermal hysteresis in resistivity of an $H_2^+$ ion-irradiated metallic $SrTiO_3$ film also implies the existence of trapped electrons together with delocalized electrons [5]. However, details of the localized state are still far from being fully understood.

In this review paper, a recent muon spin rotation ($\mu^+$SR) study on a paramagnetic defect complex formed upon implantation of $\mu^+$ pseudo-proton into $SrTiO_3$ [14] is reviewed with a specific focus on the relation with the experimental signatures of coexisting delocalized and localized electrons in the hydrogen-irradiated metallic $SrTiO_3$ films. The thermodynamic stability of the paramagnetic defect complex, which is composed of interstitial $\mu^+$ and $Ti^{3+}$ small polaron, is discussed on the basis of density functional theory (DFT) calculations in the generalized gradient approximation (GGA) $+U$ scheme for a corresponding hydrogen defect complex.

## II. MUON SPIN ROTATION EXPERIMENT
### A. Modeling isolated hydrogen with muons

Muonium (Mu) defects formed upon implantation of $\mu^+$ in insulators or semiconductors have been utilized as an experimentally accessible model of isolated hydrogen [15,16]. The electronic structures of isolated H and Mu defects are substantially identical, as can be seen when comparing atomic $Mu^0$ ($=\mu^+e^-$) and $H^0$ with respect to their reduced mass ratio ~1, where the superscript on the right of the (pseudo) element symbols indicates the total charge of corresponding atomic or defect states. Microscopic insight into Mu defects can be obtained with the $\mu^+$SR spectroscopy, which is a β-detected magnetic resonance technique using a spin-polarized $\mu^+$ beam. Technical details of $\mu^+$SR are given in Sec. II.B, which can also be found in literature, such as Ref. [17].

Characterizing a paramagnetic $Mu^0$ defect, or a paramagnetic muon-electron complex, is a relatively easy task in comparison with distinguishing two ionized Mu defect species, $Mu^+$ and $Mu^-$. Hyperfine interactions between a muon spin $I$ and an electron spin $s$ in the paramagnetic $Mu^0$ defect under a magnetic field $B$ can be modeled with the following spin Hamiltonian,

$$\mathcal{H}/h = \boldsymbol{\nu_e} \cdot \boldsymbol{s} - \boldsymbol{\nu_\mu} \cdot \boldsymbol{I} + \mathbf{A_{hf}} \boldsymbol{s} \cdot \boldsymbol{I}, \quad (1)$$
$$\mathbf{A_{hf}} = A_{iso}\mathbf{E} + \mathbf{A_{dip}},$$

where $\boldsymbol{\nu_e} = |g|\mu_B B/h$ is the electron Larmor frequency, $g(\sim 2)$ is the electron $g$-factor; $\boldsymbol{\nu_\mu} = \gamma_\mu B/2\pi$ is the muon Larmor frequency, $\gamma_\mu (= 2\pi \times 135.53$ MHz/T) is the muon gyromagnetic ratio; $A_{iso}$ is the isotropic hyperfine coupling constant due to the Fermi contact interaction, $\mathbf{E}$ is the $3 \times 3$ identity tensor; and $\mathbf{A_{dip}}$ is the dipolar hyperfine coupling tensor. In a point-dipole approximation, the dipolar part reduces to the following simple form,

$$\mathbf{A_{dip}} = A_{dip}\begin{pmatrix} -1/2 & 0 & 0 \\ 0 & -1/2 & 0 \\ 0 & 0 & 1 \end{pmatrix},$$

in the reference frame of its principal axes, where $A_{dip}$ is a function of the electron-muon distance and the magnitude of electron's effective magnetic moment. Muon spin precession frequencies, $f$s, which are quantities obtained with the $\mu^+$SR spectroscopy, can be derived from the spin Hamiltonian by diagonalizing the $\mathcal{H}/h$ matrix and taking the deference between its eigenvalue pairs corresponding to muon spin flip transitions. The electronic structure of the paramagnetic muon-electron complex is sharply reflected in the frequency shift, $\delta f = f - |\boldsymbol{\nu_\mu}|$, via the hyperfine term.

The statistical mixture of Mu defect species observed by $\mu^+$SR is generally not identical to that in global thermal equilibrium because the length of observation time after forced implantation of $\mu^+$ is restricted by its mean lifetime of $\tau_\mu = 2.2$ μs. Indeed, Mu mixtures observed at low temperatures in archetypal semiconductors, where interstitial Mu behaves as a negative-$U$ impurity, excessively contain metastable $Mu^0$ species [18]. Their temperature dependences reflect charge-state transitions toward equilibrium, which become detectable as the transition rates fall into the microsecond sensitivity range. Owing to this unique feature of μ$^+$SR, the positions of thermodynamic donor or acceptor levels for Mu defect species in bandgaps, $E(+/0)$ or $E(0/-)$, can be estimated from thermal activation behavior of corresponding charge transitions from the metastable Mu$^0$ state.

## B. Experimental setup for detecting muon spin precession in transverse magnetic fields

The positively charged muon, μ$^+$, is an unstable particle and decays into an electron neutrino, a muon antineutrino, and a positron with the mean lifetime $\tau_\mu$. The positron is preferentially emitted along the muon spin direction at the instant of its decay as a consequence of the parity violation in the weak interaction. This property makes it possible to reconstruct the motion of muon spins from the asymmetry in the positron emission.

In time-differential μ$^+$SR experiments, spin-polarized muons are implanted into a sample and the time evolution of the spin polarization $P(t)$ is monitored through the use of fast plastic scintillation detectors, where the time origin ($t = 0$) is set at the moment of muon arrival at the sample. For the purpose of observing high-frequency Larmor precession of the muon spin ensemble in magnetic fields higher than 0.1 T, a transverse-field experimental geometry is usually employed, as shown in Fig. 1. In this μ$^+$SR mode, which is available in dc muon beam facilities, such as Paul Scherrer Institut (PSI) and TRIUMF, a muon beam is incident to the sample with its spin-polarization direction ideally perpendicular to the muon transport axis and a magnetic field is applied along this axis to cause the Larmor precession. A clock is started when an incoming muon hits the thin plastic scintillator, TM, in front of the sample, then it is stopped when a decay positron is detected by any one of positron detectors arranged around the sample in a symmetric manner (U or D in the case of Fig. 1). This measurement cycle is repeated typically 10$^7$ times to build up time histograms of the decay events. For the positron detector $i$(=U, D, …), the histogram has the following form,

$$N_i(t) = N_i^0 \exp(-t/\tau_\mu)(1 + A\boldsymbol{P}(t) \cdot \boldsymbol{n}_i),$$

where $N_i^0$ is a factor that is proportional to the total dose, $A$ is effective asymmetry, and $\boldsymbol{n}_i$ is a unit vector pointing in the direction of the detector $i$ from the sample position. In the configuration of Fig. 1, the scalar projection of $\boldsymbol{P}(t)$ onto $\boldsymbol{n}_U$, $P(t)$, can be extracted from $N_U(t)$ and $N_D(t)$ through the following relation,

$$[N_U - \alpha N_D]/[N_U + \alpha N_D] = AP(t),$$

where $\alpha(= N_U^0/N_D^0)$ is the counting efficiency correction factor and $\boldsymbol{n}_D = -\boldsymbol{n}_U$ was used in the transformation. The muon spin precession frequencies, $f$s, are obtained from curve fitting analysis of $P(t)$ with a sum of damped cosine functions. For visual inspection of the frequency distribution, the time spectrum is often transformed into the frequency domain using a discrete Fourier transform routine.

## C. Mu$_i^+$-bound Ti$^{3+}$ small polaron in SrTiO$_3$

μ$^+$SR studies of SrTiO$_3$ have been performed by several groups [14,19,20]. Here, that by Ito *et al.* [14] is reviewed, focusing on the hyperfine structure of a paramagnetic Mu-related center at 1.7 K. Conventional μ$^+$SR measurements on a nominally undoped SrTiO$_3$ (001) wafer were performed at PSI, Switzerland, using the general-purpose surface-muon instrument (GPS) [21] and a spin-polarized surface μ$^+$ beam in the transverse-field (TF) configuration (Fig. 1). TFs up to 0.5 T were applied along the cubic [001] direction. The electronic structures of Mu defect species *in the dilute limit*, which formed upon implantation of μ$^+$, were investigated on the basis of μ$^+$SR frequency shifts.

Figure 2(a) shows the μ$^+$SR frequency spectrum at 1.7 K under a TF of 0.5 T. The central sharp line at the frequency of $\nu_\mu(= \gamma_\mu B/2\pi)$ corresponds to the interstitial Mu$^+$ donor state (Mu$_i^+$) without binding any unpaired electron. On the other hand, the four satellite lines are a clear signature of a paramagnetic Mu defect species, which binds an unpaired electron; the frequency shifts from the central line reflect the magnitude and anisotropy of the hyperfine interaction between μ$^+$ and the unpaired electron. The asymmetric paramagnetic shifts are typical for cases where the center of gravity of the spin density distribution is displaced from the Mu site by an atomic-scale length and consequently a dipolar contribution dominates in the hyperfine coupling. This suggests that the unpaired electron is localized primarily at a Ti site in close proximity to a $Mu_i^+$ donor to form a polaronic center [22-24], as shown in Fig. 2(b).

The simple spin Hamiltonian in eq. (1), composed of electron and muon Zeeman terms and a hyperfine coupling term, was used for numerical analysis to verify the donor-bound polaron hypothesis. The hyperfine interactions between the muon and electron spins were modelled with an isotropic contact term and a classical dipolar term within the point-dipole approximation. According to the spin Hamiltonian, the $B$ dependence of the paramagnetic shift was formulated as $\delta f(B; \theta, A_{iso}, A_{dip})$, where $\theta$ is the angle between ***B*** and an electron-to-muon vector ***r***. A global fit to $\delta f(B; \theta, A_{iso}, A_{dip})$ was performed for the inner and outer satellite pairs with shared $A_{iso}$ and $A_{dip}$. Solid curves in Fig. 2(c) are the best fit with $A_{iso}$=1.4(3) MHz, and $A_{dip}$=15.5(2) MHz, and $\theta$=33(1)° or 180 – 33(1)° for the outer pair and $\theta$=62(1)° or 180 – 62(1)° for the inner pair. The values for $\theta$ are consistent with the local structure for interstitial $H^+$ in oxides, which usually forms an O-H group with its bond length of about 1 Å, as shown in Fig. 2(b). The $A_{dip} \gg A_{iso}$ relation is also in line with the donor-bound polaron model. The magnitude of the effective magnetic moment at the adjacent Ti site was roughly estimated from $A_{dip}$ to be of order unity in $\mu_B$, which is consistent with the nominally $Ti^{3+}$ small polaron. From these observations and analysis, the satellite lines in Fig. 2(a) were ascribed to a $Mu_i^+$-bound $Ti^{3+}$ small polaron.

The height of an activation barrier between the electron bound and unbound states can be estimated from the temperature dependence of signal intensities for corresponding Mu defect species. Figure 3(a) shows the temperature evolution of $\mu^+$SR frequency spectra at $B$ = 0.5 T. The intensity for the electron bound state (satellite lines) gradually decreases and is transferred to that for the unbound state (central line) with increasing temperature. From this behavior, the activation energy $E_a$ for thermal dissociation of the $Mu_i^+$-bound small polaron was determined to be 30(3) meV.

While it seems unusually small for the energy that separates the small polaron bound state and the unbound state, this situation can be understood by considering the energy balance for polaron formation in the presence of lattice distortion caused by the localized electron, as shown in Fig. 3(b). The $E_a$ obtained from $\mu^+$SR corresponds to a thermal activation energy for transferring the trapped electron to the local conduction band. The small $E_a$ suggests that the absolute value of polaron formation energy, $|E_{pol}|$, is also small as a consequence of the energy balance between the benefit in the electron system, $E_{el}$, and the loss in the lattice system, $E_{st}$. It should be noted that it is not obvious only from $\mu^+$SR whether $E_{pol}$ is positive or not. For clarifying this, DFT studies on the stability of $H_i^+$-bound $Ti^{3+}$ small polaron are described in the next section. The vertical excitation energy $E_{el}$ can be associated with the deep in-gap state detected by photoemission spectroscopy in the hydrogenated $SrTiO_3$ film within the time scale where lattice atoms are substantially frozen [13]. The electron in the local conduction band is finally transferred to the electron reservoir if this state is lower in energy than the other defect configurations.

For taking the discussion further, attention must be paid to the non-equilibrium nature of Mu mixture just after $\mu^+$ implantation, as mentioned in Sec. IIA. At low temperatures, as-implanted non-equilibrium distribution of Mu species, which excessively involves the metastable paramagnetic species, is maintained over the $\mu^+$SR time window. The transition to the stable diamagnetic state gradually becomes visible with increasing temperature as the transition rate falls into the microsecond sensitivity range (Fig. 3(a)). A similar situation could arise in the hydrogenated $SrTiO_3$ film prepared by $H_2^+$ ion irradiation at low temperature. Indeed, thermal hysteresis in resistivity of this film implies that the as-implanted mixture of H states excessively contains neutral species [5]. This neutral state is suggested to be an $H_i^+$-bound small polaron from the similarity to the $\mu^+$ implantation experiment.

## III. DFT CALCULATIONS

## A. Computational details

All calculations were performed within the DFT framework with a Hubbard $U$ correction by using the QUANTUM ESPRESSO code [25,26]. The generalized gradient approximation (GGA) using Perdew-Burke-Ernzerhof (PBE) exchange correlation functional was adopted. PAW-type pseudopotentials with H(1$s$), Sr(4$s$, 4$p$, 5$s$), Ti(3$s$, 3$p$, 4$s$, 3$d$) and O(2$s$, 2$p$) valence states were used. Wave functions were expanded in plane waves with a cutoff of 70 Ry for the kinetic energy and 600 Ry for the charge density.

The SrTiO$_3$ structure was described using a 3×3×3 supercell of the 5-atom cubic primitive cell. Brillouin zone sampling with a Monkhorst-Pack $k$-point mesh of 3×3×3 was applied. The finer $k$-point mesh of 5×5×5 was used for plotting density of states (DOS). The Hubbard $U$ potential on Ti 3$d$ states, $U_{Ti}$, was set to 4.74 eV according to Ref. [27]. Gaussian smearing with a broadening of 0.01 Ry was used for structure optimization calculations, where atomic positions were relaxed with keeping the lattice constant fixed at the optimized value for stoichiometric SrTiO$_3$. In all cases, atomic forces were converged to within 7.7×10$^{-3}$ eV/Å. Then, the self-consistent field calculation for the optimized structure was refined with the optimized tetrahedron method [28] for avoiding artifacts associated with the Gaussian smearing. The structure optimization calculations were started from spin-unpolarized initial states unless otherwise specified.

The formation energy of an interstitial H defect in a charge state $q$, $E_f(\text{H}^q)$, was calculated as described in Ref. [9,29]:

$E_f(\text{H}^q) = E_t(\text{H}^q) - E_t(\text{host}^0) - 1/2\mu_{\text{H}_2} + q(\mu_{\text{VBM}} + E_F) + E_{corr}$,

where $E_t(\text{H}^q)$ is the total energy of a defective supercell in a charge state $q$, $E_t(\text{host}^0)$ is the total energy of a charge neutral host supercell, $\mu_{\text{H}_2}$ is the chemical potential of an H$_2$ molecule, $\mu_{\text{VBM}}$ is the chemical potential of the electron reservoir at the valence-band maximum (VBM), and $E_F$ is the Fermi energy with reference to $\mu_{\text{VBM}}$. $E_{corr}$ is a small potential-alignment term for charged defective supercells, which was obtained in accordance with a standard procedure [30].

## B. Results and discussion

Figure 4(a) shows the defect formation energies for $q = -1$, 0, and +1 as functions of $E_F$. The three lines except for the lower horizontal line were obtained from calculations starting from spin-unpolarized initial states. These lines cross at an $E(+/0/-)$ transition level in the conduction band, which is consistent with previous DFT studies on cubic SrTiO$_3$ and BaTiO$_3$ [9,29]. The local structures of these defects are nearly identical; the interstitial H is in an ionized state with a formal charge +1 and creates an O-H bond with an oxygen ion in the host lattice, as shown in Fig. 4(c). Excess electrons in the supercells with $q = 0$ or $-1$ simply fill in the bottom of the conduction band in a delocalized manner, as shown in Fig. 4(d).

In addition to the delocalized solutions, a localized solution, which corresponds to the lower horizontal line in Fig. 4(a), was also obtained for the charge neutral supercell ($q = 0$) from a calculation starting from a spin-polarized initial state with a magnetic moment of 1 $\mu_B$ set at an adjacent Ti ion. The local structure of the defect was shown in Fig. 4(e) with a spin density isosurface. This indicates that an unpaired electron is localized predominantly at a single Ti ion to form an H$_i^+$-Ti$^{3+}$ small polaron complex. The DOS plot shown in Fig. 4(f) is in line with the small polaron picture, having a deep single particle state in the bandgap. On the other hand, the defect formation energy diagram for $U_{Ti}$ = 4.74 eV (Fig. 4(a)) indicates that a thermodynamic donor level, $E(+/0)$, can form just below the conduction band minimum, which is associated with electron transfer from the H$_i^+$-Ti$^{3+}$ complex to the conduction band. These features qualitatively agree with the polaron formation mechanism depicted in Fig. 3(b).

The present DFT calculations are not conclusive as to whether the $E(+/0)$ donor level forms in the bandgap or not because the relative position of the $E(+/0)$ level with respect to the conduction band minimum varies with the $U_{Ti}$ parameter, as shown in Fig. 4(b). On the other hand, this figure together with Fig. 4(a) also suggests that the H$^+$ state and the H$^0$ state with a localized electron are nearly degenerated in $n$-type carrier rich conditions with $E_F$ close to the conduction band minimum for a realistic range of $U_{Ti}$, regardless of the sign of $E_g - E(+/0)$. The experimental signature of coexisting localized and delocalized electrons detected by photoemission spectroscopy in a hydrogenated metallic $SrTiO_3$ film [13] may be understood on the basis of this pseudo-degeneracy.

## IV. CONCLUSIONS

The $\mu^+$SR study on the $Mu_i^+$-$Ti^{3+}$ defect complex formed upon implantation of $\mu^+$ into $SrTiO_3$ [14] was reviewed with a specific focus on the relation with the experimental signatures of coexisting delocalized and localized electrons in hydrogen-irradiated metallic $SrTiO_3$ films with the support of DFT calculations. It was shown through this review that the donor-bound polaron concept can lead to a coherent understanding of seemingly puzzling experimental results in hydrogenated films. It was also mentioned that $\mu^+$SR is useful for simulating non-equilibrium phenomena in hydrogen-irradiated systems. The importance of this technique is expected to increase further as the application of the H implantation methods expands.

### Acknowledgments

Discussions and correspondence are gratefully acknowledged with W. Higemoto, A. Koda, K. Shimomura, K. Nishiyama, K. Fukutani, Y. Iwazaki, S. Tsuneyuki, R. Kadono, and N. Nishida. The DFT calculations were conducted with the supercomputer HPE SGI8600 in Japan Atomic Energy Agency. This research was partially supported by JSPS KAKENHI Grant (No. 21H05102, 20K12484, 20H01864, and 20H02037).

## References

[1] J. F. Schooley, W. R. Hosler, and M. L. Cohen, Phys. Rev. Lett. **12**, 474 (1964).
https://doi.org/10.1103/PhysRevLett.12.474

[2] H. P. R. Frederikse and W. R. Hosler, Phys. Rev. **161**, 822 (1967).
https://doi.org/10.1103/PhysRev.161.822

[3] G. Herranz, M. Basletić, O. Copie, M. Bibes, A. N. Khodan, C. Carrétéro, E. Tafra, E. Jacquet, K. Bouzehouane, A. Hamzić, and A. Barthélémy, and G. Herranz, Appl. Phys. Lett. **94**, 012113 (2009).
https://doi.org/10.1063/1.3063026

[4] A. Spinelli, M. A. Torija, C. Liu, C. Jan, and C. Leighton Phys. Rev. B **81**, 155110 (2010).
https://doi.org/10.1103/PhysRevB.81.155110

[5] R. Nakayama, M. Maesato, T. Yamamoto, H. Kageyama, T. Terashima, and H. Kitagawa, Chem. Commun. **54**, 12439 (2018).
https://doi.org/10.1039/C8CC07021K

[6] C. Cen, S. Thiel, J. Mannhart, and J. Levy, Science **323**, 1026 (2009)
https://doi.org/10.1126/science.1168294

[7] S. Klauer and M. Wöhlecke, Phys. Rev. Lett. **68**, 3212 (1992.)
https://doi.org/10.1103/PhysRevLett.68.3212

[8] G. Weber, S. Kapphan, and M. Wöhlecke, Phys. Rev. B **34**, 8406 (1986).
https://doi.org/10.1103/PhysRevB.34.8406

[9] Y. Iwazaki, Y. Gohda, and S. Tsuneyuki, APL Mater. **2**, 012103 (2014).
https://doi.org/10.1063/1.4854355

[10] Y. Kobayashi, O. J. Hernandez, T. Sakaguchi, T. Yajima, T. Roisnel, Y. Tsujimoto, M. Morita, Y. Noda, Y. Mogami, A. Kitada, M. Ohkura, S. Hosokawa, Z. Li, K. Hayashi, Y. Kusano, J. eun Kim, N. Tsuji, A. Fujiwara, Y. Matsushita, K. Yoshimura, K. Takegoshi, M. Inoue, M. Takano and H. Kageyama, Nature Mater **11**, 507 (2012).
https://doi.org/10.1038/nmat3302

[11] T. Sakaguchi, Y. Kobayashi, T. Yajima, M. Ohkura, C. Tassel, F. Takeiri, S. Mitsuoka, H. Ohkubo, T. Yamamoto, J. eun Kim, N. Tsuji, A. Fujihara, Y. Matsushita, J. Hester, M. Avdeev, K. Ohoyama, and H. Kageyama, Inorg. Chem. **51**, 11371 (2012).
https://doi.org/10.1021/ic300859n

[12] T. U. Ito, A. Koda, K. Shimomura, W. Higemoto, T. Matsuzaki, Y. Kobayashi, and H. Kageyama, Phys. Rev. B **95**, 020301(R) (2017).
https://doi.org/10.1103/PhysRevB.95.020301

[13] M. D'Angelo, R. Yukawa, K. Ozawa, S. Yamamoto, T. Hirahara, S. Hasegawa, M. G. Silly, F. Sirotti, and I. Matsuda, Phys. Rev. Lett. **108**, 116802 (2012).
https://doi.org/10.1103/PhysRevLett.108.116802

[14] T. U. Ito, W. Higemoto, A. Koda, and K. Shimomura, Appl. Phys. Lett., **115**, 192103 (2019).

https://doi.org/10.1063/1.5125919

[15] S. F. J. Cox, Rep. Prog. Phys. **72**, 116501 (2009).

https://doi.org/10.1088/0034-4885/72/11/116501

[16] T. U. Ito, W. Higemoto, and K. Shimomura, J. Phys. Soc. Jpn. **89**, 051007 (2020).

https://doi.org/10.7566/JPSJ.89.051007

[17] A. Yaouanc and P. Dalmas de Réotier, *Muon Spin Rotation, Relaxation, and Resonance: Applications to Condensed Matter* (Oxford University Press, Oxford, U.K., 2010) 1st ed.

[18] R. L. Lichti, K. H. Chow, and S. F. J. Cox, Phys. Rev. Lett. **101**, 136403 (2008).

https://doi.org/10.1103/PhysRevLett.101.136403

[19] Z. Salman, T. Prokscha, A. Amato, E. Morenzoni, R. Scheuermann, K. Sedlak, and A. Suter, Phys. Rev. Lett. **113**, 156801 (2014).

https://doi.org/10.1103/PhysRevLett.113.156801

[20] D. P. Spencer, D. G. Fleming and J. H. Brewer, Hyperfine Interact. **18**, 567 (1984).

https://doi.org/10.1007/BF02064869

[21] A. Amato, H. Luetkens, K. Sedlak, A. Stoykov, R. Scheuermann, M. Elender, A. Raselli, and D. Graf, Rev. Sci. Instrum. **88**, 093301 (2017).

https://doi.org/10.1063/1.4986045

[22] S. F. J. Cox, J. L. Gavartin, J. S. Lord, S. P. Cottrell, J. M. Gil, H. V. Alberto, J. Piroto Duarte, R. C. Vilão, N. Ayres de Campos, D. J. Keeble, E. A. Davis, M. Charlton, and D. P. van der Werf, J. Phys.: Condens. Matter **18**, 1079 (2006).

http://dx.doi.org/10.1088/0953-8984/18/3/022

[23] K. Shimomura, R. Kadono, A. Koda, K. Nishiyama, and M. Mihara, Phys. Rev. B **92**, 075203 (2015).

https://doi.org/10.1103/PhysRevB.92.075203

[24] R. C. Vilão, R. B. L. Vieira, H. V. Alberto, J. M. Gil, A. Weidinger, R. L. Lichti, B. B. Baker, P. W. Mengyan, and J. S. Lord, Phys. Rev. B **92**, 081202(R) (2015).

https://doi.org/10.1103/PhysRevB.92.081202

[25] P. Giannozzi, S. Baroni, N. Bonini, M. Calandra, R. Car, C. Cavazzoni, D. Ceresoli, G. L. Chiarotti, M. Cococcioni, I. Dabo, A. Dal Corso, S. de Gironcoli, S. Fabris, G. Fratesi, R. Gebauer, U. Gerstmann, C. Gougoussis, A. Kokalj, M. Lazzeri, L. Martin-Samos, N. Marzari, F. Mauri, R. Mazzarello, S. Paolini, A. Pasquarello, L. Paulatto, C. Sbraccia1, S. Scandolo, G. Sclauzero, A. P. Seitsonen, A. Smogunov, P. Umari and R. M. Wentzcovitch, J. Phys.: Condens. Matter **21** 395502 (2009).

http://dx.doi.org/10.1088/0953-8984/21/39/395502

[26] P. Giannozzi, O. Andreussi, T. Brumme, O. Bunau, M. Buongiorno Nardelli, M. Calandra, R. Car, C. Cavazzoni, D. Ceresoli, M. Cococcioni, N. Colonna, I. Carnimeo, A. Dal Corso, S. de Gironcoli, P. Delugas, R. A. Distasio, A. Ferretti, A. Floris, G. Fratesi, G. Fugallo, R. Gebauer, U. Gerstmann, F. Giustino, T. Gorni, J. Jia, M. Kawamura, H. Y. Ko, A. Kokalj, E. Kücükbenli, M. Lazzeri, M. Marsili, N. Marzari, F. Mauri, N. L. Nguyen, H. V. Nguyen, A. Otero-De-La-Roza, L. Paulatto, S. Poncé, D. Rocca, R. Sabatini, B. Santra, M. Schlipf, A. P. Seitsonen, A. Smogunov, I. Timrov, T. Thonhauser, P. Umari, N. Vast, X. Wu, and S. Baroni, J. Phys.: Condens. Matter **29**, 465901 (2017).

https://doi.org/10.1088/1361-648X/aa8f79

[27] C. Ricca, I. Timrov, M. Cococcioni, N. Marzari, and U. Aschauer, Phys. Rev. Research **2**, 023313 (2020).

https://doi.org/10.1103/PhysRevResearch.2.023313

[28] M. Kawamura, Y. Gohda, and S. Tsuneyuki, Phys. Rev. B **89**, 094515 (2014).

https://doi.org/10.1103/PhysRevB.89.094515

[29] Y. Iwazaki, T. Suzuki, and S. Tsuneyuki, J. Appl. Phys. **108**, 083705 (2010).

http://dx.doi.org/10.1063/1.3483243

[30] C. Persson, Y.-J. Zhao, S. Lany, and A. Zunger, Phys. Rev. B **72**, 035211 (2005).

https://doi.org/10.1103/PhysRevB.72.035211

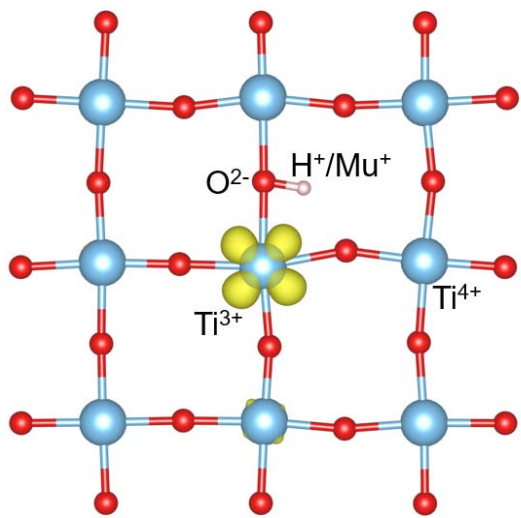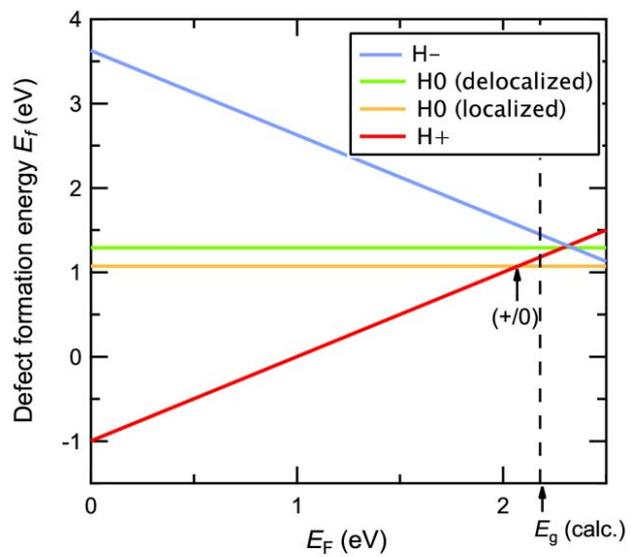

Graphical abstract

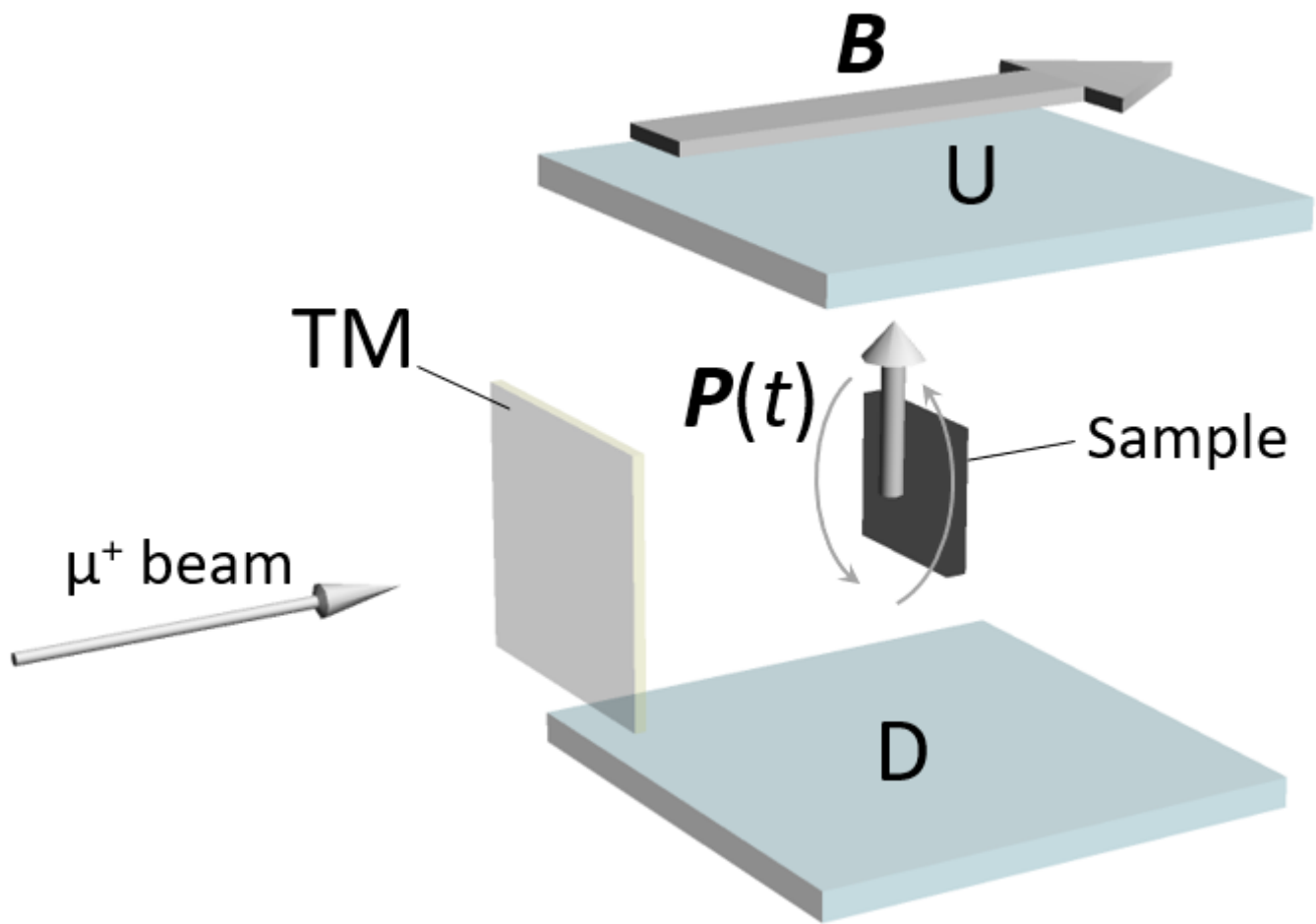

Figure 1: Schematic illustration of the transverse-field experimental geometry.

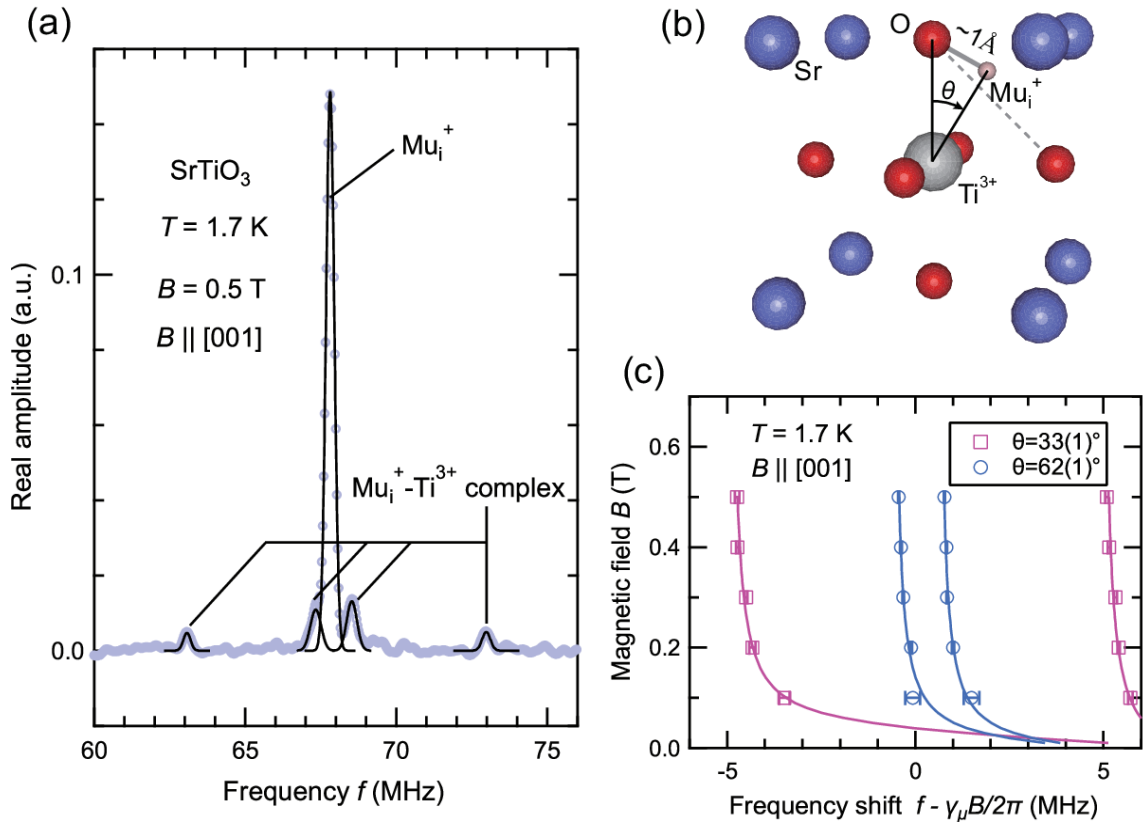

Figure 2: (a) μ$^+$SR frequency spectrum at 1.7 K under a TF of 0.5 T applied along the cubic [001] direction. (b) Schematic illustration of the Mu$_i^+$-bound Ti$^{3+}$ small polaron in SrTiO$_3$. (c) TF dependence of frequency shifts for two pairs of paramagnetic lines. The solid curves represent the best fits to $\delta f(B; \theta, A_{iso}, A_{dip})$.

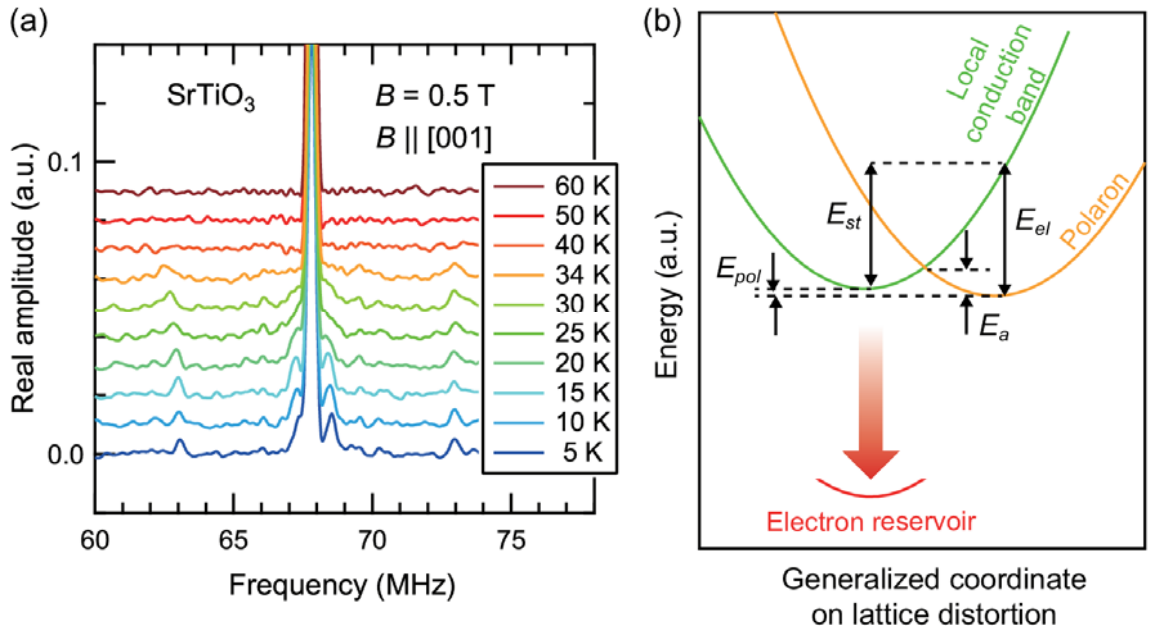

Figure 3: (a) Temperature evolution of μ⁺SR frequency spectra in a TF of 0.5 T applied along the cubic [001] direction. (b) Schematic illustration of the energy balance for polaron formation in the presence of lattice distortion.

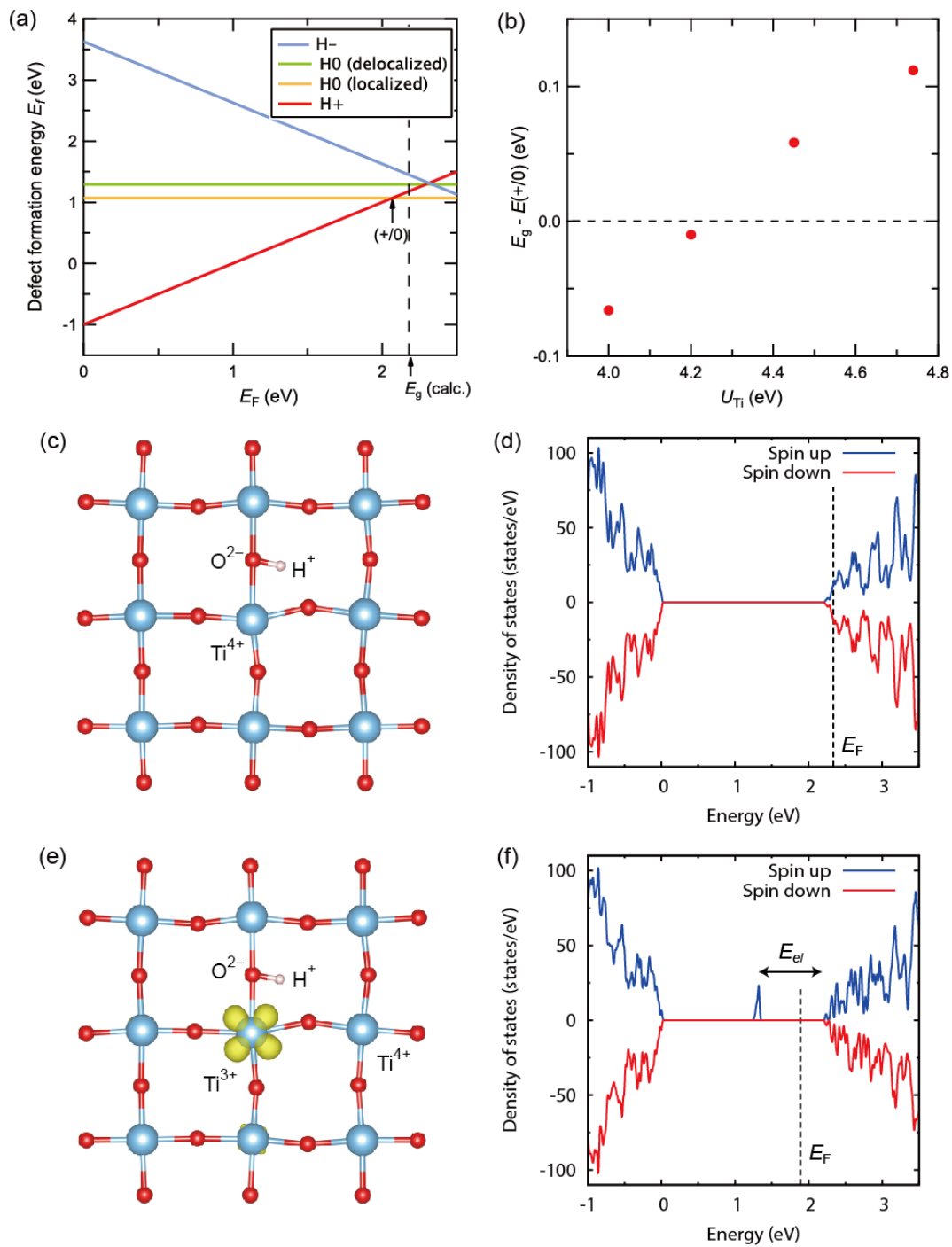

Figure 4: (a) Defect formation energies for $q = -1, 0$, and $+1$ with $U_{Ti} = 4.74$ eV as functions of $E_F$. The calculated bandgap is smaller than experimental one, which is ascribed to well-known shortcomings of GGA-PBE calculations. (b) Depth of the $E(+/0)$ level below the conduction band minimum as a function of $U_{Ti}$. (c) Local defect structure corresponding to the $H^0$ delocalized solution. (d) DOS for the $H^0$ delocalized solution. (e) Local defect structure corresponding to the $H^0$ localized solution shown with a spin density isosurface. (f) DOS for the $H^0$ localized solution.